\documentclass[twocolumn,english,prl,floatfix,showpacs]{revtex4-1}
\usepackage[T1]{fontenc}
\usepackage[latin9]{inputenc}
\usepackage{textcomp}
\usepackage{amsmath}
\usepackage{amssymb}
\usepackage{graphicx}
\PassOptionsToPackage{normalem}{ulem}
\usepackage{ulem}

\makeatletter

\providecommand{\tabularnewline}{\\}

\graphicspath{ {./figures/} }

\usepackage{babel}

\usepackage{babel}

\makeatother

\usepackage{babel}
\begin{document}
\title{Analysis of the Relation between Quadratic Unconstrained Binary Optimization
(QUBO) and the Spin Glass Ground-State Problem}
\author{Stefan Boettcher}
\affiliation{Department of Physics, Emory University, Atlanta, GA 30322; USA}
\begin{abstract}
We analyze the transformation of QUBO from its conventional Boolean
presentation into an equivalent spin glass problem with coupled $\pm1$
spin variables exposed to a site-dependent external field. We find
that in a widely used testbed for QUBO these fields tend to be rather
large compared to the typical coupling and many spins in each optimal
configuration simply align with the fields irrespective of their constraints.
Thereby, the testbed instances tend to exhibit large redundancies
- seemingly independent variables which contribute little to the hardness
of the problem, however. We demonstrate various consequences of this
insight, for QUBO solvers as well as for heuristics developed for
finding spin glass ground states. To this end, we implement the Extremal
Optimization (EO) heuristic, in a new adaptation for the QUBO problem.
We also propose a novel way to assess the quality of heuristics for
increasing problem sizes based on asymptotic scaling.
\end{abstract}
\maketitle

\section{Introduction\label{sec:Intro}}

Quadratic unconstrained binary optimization (QUBO) is a versatile
NP-hard combinatorial problem with applications in operations research~\cite{Lu10}
and financial assets management, for example. It has recently been
studied also as a benchmark challenge for the D-Wave quantum annealer~\cite{McGeoch13}
or for a new generation of classical optimizers based on GPU-technology~\cite{Aramon2019}.
Cutting-edge classical algorithms for QUBO, developed in the engineering
literature, are based on TABU search~\cite{Wang13,Glover10,Lu10,Boros07,Palubeckis06}
and a variety of other heuristics~\cite{Kochenberger2014}. From
a statistical physics perspective, these developments are tantalizing
for the fact that the generic formulation of QUBO appears to be identical
to that of the Ising spin-glass Hamiltonian. While this connection
has long be realized~\cite{Barahona88}, it poses a conundrum that
has not be commented on previously, and whose resolution could be
of importance for both, the study of the low-energy structure of spin
glasses as well as the understanding of its combinatorial hardness,
for example, to assess the capabilities of the aforementioned solvers,
classical and quantum.

Short of a real quantum computing solution, our only hope to find
approximate solutions of reasonable quality for large-size instances
of many NP-hard combinatorial optimization problems stems from the
design of heuristic methods~\cite{Dagstuhl04,Hoos04,Osman96,Voss99}.
From that perspective, it is somewhat surprising to find that seemingly
equivalent instances of QUBO are routinely solved with well up to
$N\approx10^{4}$ variables~\cite{Glover10,Lu10,Palubeckis06,Wang13}
while solvers for comparable spin glasses already struggle with instances
of $N\approx10^{3}$ variables to converge without incurring unacceptable
systematic errors~\cite{hartmann:01b,Palassini99,Pal06b,Boettcher10b,EOSK}.
Could adapting those highly developed QUBO solvers from the operations
research literature provide a significant new inroad into investigations
of spin glasses? What we find instead, unfortunately, is that there
is an inherent weakness in the definition of the typical testbeds
employed to assess QUBO solvers, which is revealed when these testbed
instances are expressed as mean-field spin glasses. Exploiting this
weakness, we apply a novel implementation of the Extremal Optimization
(EO) heuristic~\cite{Boettcher00,Boettcher01a,Dagstuhl04,Middleton04}
to the QUBO problem that performs on par with QUBO solvers for such
large instances. In turn, we demonstrate that a naive application
of a typical QUBO solver performs  poorly for the spin glass. However,
it would be of considerable physics interest to harness the power
of TABU search and have experts in the design of QUBO solvers tune
their implementations for spin glass problems for a fair comparison. 

Besides of the caution against over-interpreting the significance
of solving ``large'' instances, our study also produces a number
of positive results. Our new implementation of EO not only serves
as an alternative QUBO solver, but its design also provides insights
that will advance the future exploration of the low-energy landscape
of Ising spin glasses in the presence of external fields. Furthermore,
we propose a powerful test for heuristic solvers that, in contrast
with traditional testbed instances, unambiguously reveals the scalability
of solvers asymptotically with problem size. 

This paper is organized as follows: in Sec.~\ref{sec:RelationSG_QUBO},
we revisit the well-known relation between QUBO and spin glasses,
with the added twist of a gauge transformation. In Sec.~\ref{sec:Using-QUBO-Solvers},
we adapt a sophisticated implementation of TABU search to study ground
states of mean-field (Sherrington-Kirkpatrick) spin glasses. In Sec.~\ref{sec:Using-EO-to},
we employ EO to study the QUBO problem in a manner that incorporates
well-known testbeds while also arguing for a novel way of quantifying
the scalability of heuristics. In Sec.~\ref{sec:Conclusions}, we
conclude with an assessment of the state of the art for solving QUBO
problems with heuristics and provide an outlook on future work.

\section{Relation Between Spin Glasses and QUBO\label{sec:RelationSG_QUBO}}

Disordered Ising spin systems in the mean-field limit have been investigated
extensively as models of combinatorial optimization problems~\cite{MPV,Percus06}.
Particularly simple are such models on (sparse) $\alpha$-regular
random graphs (``Bethe-lattice''), where each vertex possesses a
fixed number $\alpha$ of bonds to randomly selected other vertices~\cite{mezard:01,Boettcher03a},
or on a (dense) fully connected graph, referred to as the Sherrington-Kirkpatrick
model (SK)~\cite{Sherrington75,MPV,EOSK}. Instances in an ensemble
are formed via a matrix $J_{ij}$ of bonds between adjacent vertices
$i$ and $j$, typically drawn randomly from a symmetric distribution
such as ${\cal N}(0,1)$ (normal, Gaussian) or $\pm1$ (bi-modal).
(Accordingly, it is $J_{ii}\equiv0$, as there are no ``self-bonds''.)
A dynamic variable $\sigma_{i}\in\{-1,+1\}$ (``spin'') is assigned
to each vertex. Interconnecting loops of existing bonds lead to competing
constraints and ``frustration''~\cite{Toulouse77}, making optimal
(minimal energy) spin configurations hard to find. In addition, we
will allow for each spin to experience an external torque due to local
magnetic fields $h_{i}$, which may also be drawn randomly or be of
uniform fixed value. In the SK problem discussed here, we will merely
consider the case of field-free instances ($h_{i}\equiv0$). However,
in the discussion of the relation between QUBO and SK, we will have
to provide for the possibility of non-zero fields. Hence, as our cost
function of this generalized problem, we endeavor to minimize the
energy $H$ of the system, 
\begin{eqnarray}
H & = & -\sum_{i=1}^{N}\sum_{j=i+1}^{N}J_{ij}\sigma_{i}\sigma_{j}-\sum_{i=1}^{N}h_{i}\sigma_{i},\label{eq:Heq}
\end{eqnarray}
over the variables $\sigma_{i}$. 

In turn, for QUBO we  minimize the cost function\footnote{In the operations research literature, QUBO is usually defined as
a maximization problem for $E$ without the sign; the conversion is
trivial.} 
\begin{equation}
E=-\sum_{i=1}^{N}\sum_{j=1}^{N}q_{ij}x_{i}x_{j},\label{eq:QUBO}
\end{equation}
over a set of $N$ Boolean variables $x_{i}\in\left\{ 0,1\right\} $.
Note that in this case it is $q_{ii}\not=0$, unlike for spin-glass
couplings in Eq.~(\ref{eq:Heq}). A generalized form of the QUBO
cost with a term linear in the variables, similar to SK with an external
field in Eq.~(\ref{eq:Heq}), is not necessary, since we can use
the identity $x_{i}\equiv x_{i}^{2}$, valid for $x_{i}\in\left\{ 0,1\right\} $,
to write any linear terms as $cx_{i}=cx_{i}^{2}$ and add weights
$c$ to that on the diagonal, $q_{ii}$. The test instances often
considered for QUBO are created by choosing symmetric weights $q_{ij}$,
drawn from a uniform (typically flat) distribution of zero mean, such
as $-100<q_{ij}<100$ filling $N\times N$ matrices with 10-100\%
density~\cite{Beasley98,Lu10,Glover10}. (It seems that samples of
sparse instances comparable to Bethe-lattices have not yet been discussed
for QUBO.) In that literature, there is a distinct focus on specific
testbeds of a few instances that are referenced for every new method
applied to the problem, in an attempt to facilitate comparisons between
the methods. Here, we merely consider a set of 10 such testbed instances
from each of the sets ``bqp1000'' and ``bqp2500'', of size $N=1000$
and 2500, respectively, to also allow for such a comparison. However,
as we will see, significant insight, especially about the scaling
with $N$ of each problem, can be gained by instead taking an ensemble
perspective, i.e., we will make cost averages obtained over a larger
number of instances taken at random from the ensemble at various sizes
$N$. 

Both problem statements, Eqs.~(\ref{eq:Heq}-\ref{eq:QUBO}), appear
to be rather similar, including the symmetric distribution of weights
and variables of a binary type, and one may wonder whether a detailed
comparison between QUBO and SK as distinct optimization problems is
warranted. Yet, the fact that spin glasses are defined for Ising variables,
$\sigma_{i}\in\left\{ \pm1\right\} $, while QUBO has Boolean variables,
$x_{i}\in\left\{ 0,1\right\} $, proves quite consequential. 

\subsection{Spin Glass as a QUBO Problem\label{subsec:Spin-Glass-as}}

For using a QUBO solver to optimize the SK spin glass problem in Sec.~\ref{sec:Using-QUBO-Solvers},
we have to rewrite the spin-glass cost function in Eq.~(\ref{eq:Heq})
in terms of the Boolean variables a QUBO solver operates on. To that
end, we assume given bonds $J_{ij}$ and fields $h_{i}$ and set $\sigma_{i}=2x_{i}-1$
to obtain
\begin{equation}
H=2E+C,\label{eq:H_QUBO}
\end{equation}
with $C=-\sum_{i=1}^{N}\sum_{j=i+1}^{N}J_{ij}+\sum_{i=1}^{N}h_{i}$
as some fixed constant for each instance. Now, $E$ takes on exactly
the form of Eq.~(\ref{eq:QUBO}) but with weights
\begin{eqnarray}
q_{ij} & = & \begin{cases}
J_{ij}, & i\not=j,\\
\\
h_{i}-\sum_{l=1}^{N}J_{il}, & i=j.
\end{cases}\label{eq:qij_SK}
\end{eqnarray}
Thus, by solving the QUBO problem for $E$ with these weights, we
easily extract the spin glass ground state $H$ via Eq.~(\ref{eq:H_QUBO}).
Note that although all $q_{ij}$ for $i\not=j$ are still simply random
numbers drawn from a symmetric distribution, the diagonal elements
$q_{ii}$ instead become \emph{extensive} sums of such numbers, unless
all $h_{i}\not=0$ and are specifically chosen as counterbalance.
Such $q_{ii}$ are still symmetrically -- but far more broadly --
distributed (by a factor $\sim\sqrt{N}$) and always determined such
that each row-sum and column-sum vanishes. Since $x_{i}^{2}\equiv x_{i}$,
those diagonal elements are apparently equivalent to a linear term
supplementing the QUBO cost-function, Eq. (\ref{eq:QUBO}). However,
the properties of such a term are quite different from the magnetic
field term in Eq. (\ref{eq:Heq}), as we will discuss in Sec. \ref{subsec:Magnatization-of-QUBO}.

\subsection{QUBO Problem as a Spin Glass\label{subsec:QUBO-Problem-as}}

Using a spin-glass solver to optimize the QUBO problem in Sec.~\ref{sec:Using-EO-to},
correspondingly, we take the QUBO weights $q_{ij}$ as given and rewrite
the QUBO variables $x_{i}$ as spins $\sigma_{i}\in\left\{ \pm1\right\} $
via $x_{i}=\frac{1}{2}\left(1+\sigma_{i}\right)$. With that, in full
analogy with Eq.~(\ref{eq:H_QUBO}), we find 
\begin{equation}
E=\frac{1}{2}H-\frac{1}{2}C\label{eq:E_SK}
\end{equation}
with a Hamiltonian as given in Eq.~(\ref{eq:Heq}) when using the
bonds and fields as
\begin{align}
h_{i} & =\sum_{j=1}^{N}q_{ij},\label{eq:Jh_QUBO}\\
J_{ij} & =\begin{cases}
q_{ij}, & i\not=j,\\
\\
0, & i=j.
\end{cases}
\end{align}
Here, $C=\sum_{i=1}^{N}\sum_{j=i+1}^{N}q_{ij}+\sum_{i=1}^{N}q_{ii}$
again is an inert constant that is easily evaluated for each instance.
Note that each single field $h_{i}$ itself becomes a symmetrically
distributed random variable of width $\sim\sqrt{N}$, a sum over an
entire row of the $q_{ij}$-matrix, if $q_{ij}$ is such a random
variable of unit width. Such a strong biasing field, as we will argue
in detail below, poses a serious problem for the design of truly hard
QUBO instances. We will discuss in more detail how to find approximate
ground states of such a spin-glass Hamiltonian with an external field
in Sec.~\ref{sec:Using-EO-to}. However, given that, the cost for
the QUBO problem follows simply from Eq.~(\ref{eq:E_SK}). 

\subsection{Gauge Transformation\label{sub:Gauge-Transformation}}

While the existence of a relation between QUBO and spin glasses is
not a novel observation~\cite{Barahona88,Kochenberger2014}, the
following consideration, albeit simple, allows for a pertinent insight
into the nature of optimal configurations of QUBO that seems to have
escaped prior notice. In general, a spin-glass Hamiltonian as in Eq.~(\ref{eq:Heq})
retains all its spectral properties (here, in particular, its ground-state
energy) under the transformation 
\begin{equation}
\sigma_{i}\to\sigma_{i}^{\prime}=\xi_{i}\sigma_{i},\qquad\xi_{i}=\pm1,\label{eq:gaugetransform}
\end{equation}
for all $i$. Then, 
\begin{eqnarray}
H\left(\left\{ \sigma_{i}^{\prime}\right\} \right) & = & -\sum_{i}\sum_{j}\xi_{i}\xi_{j}J_{ij}\sigma_{i}\sigma_{j}-\sum_{i}\xi_{i}h_{i}\sigma_{i},\\
 & = & -\sum_{i}\sum_{j}J_{ij}^{\prime}\sigma_{i}\sigma_{j}-\sum_{i}h_{i}^{\prime}\sigma_{i},\nonumber 
\end{eqnarray}
when we identify
\begin{equation}
J_{ij}^{\prime}=\xi_{i}\xi_{j}J_{ij},\qquad h_{i}^{\prime}=\xi_{i}h_{i},\label{eq:gauge}
\end{equation}
Thus, the transformation in Eq. (\ref{eq:gaugetransform}) leaves
the spin-glass Hamiltonian invariant. We note that such an invariance
does not exist for the (Boolean) QUBO problem, as the corresponding
transformation $x_{i}\to x_{i}^{\prime}=1-x_{i}$ on only select sites
$i$ modifies the QUBO Hamiltonian in Eq. (\ref{eq:QUBO}). 

Via Eq.~(\ref{eq:gauge}), we are now free to ``gauge'' our spin
variables in any form desirable. For our purposes, it is enlightening
here to choose the set $\left\{ \xi_{i}\right\} $ such that \emph{all}
external fields are \emph{non-negative}, $h_{i}^{\prime}\geq0$ for
all $i$ in Eq.~(\ref{eq:gauge}). We can easily obtain the solution
of the original problem via $H(\left\{ \xi_{i}\sigma_{i}\right\} )=H^{\prime}(\left\{ \sigma_{i}\right\} )$,
in particular, for the optimal configuration. It is now intuitive
to ask: To what extend do spins in the optimal configuration align
with their external field, irrespective of the mutual couplings $J_{ij}$?
We will address that question in Sec.~\ref{sec:Using-EO-to}. First,
we will explore how a QUBO solver fares in finding SK ground states.

\section{Using QUBO Solvers for SK\label{sec:Using-QUBO-Solvers}}

Here, we will apply a freely available QUBO solver, namely the Iterated
Tabu Search (ITS) designed by G. Palubeckis in the implementation
download from https://www.personalas.ktu.lt/\textasciitilde ginpalu/.
In Ref.~\cite{Palubeckis06}, this implementation of ITS was used
to reproduce the best-known results for various QUBO testbed instances
(such as those discussed in Sec.~\ref{sec:Using-EO-to}) of up to
$N=7000$ variables. Similar results were found with other implementations
of Tabu-based QUBO solvers~\cite{Kochenberger2014}, and we assume
the following observations to be generic for that class of solvers.
We modify the ITS implementation only in so far as to input a large
number of instances drawn from the SK-ensemble with bimodal bonds
and to convert those into QUBO, as introduced in Sec.~\ref{subsec:Spin-Glass-as}.
Experts in Tabu Search will note that no effort has been undertaken
to tune the heuristic for the different ensemble, for which we have
insufficient experience to accomplish. Thus, the following results
are meant to serve as an illustration that a successful application
to large QUBO instances does not imply the same for spin glasses. 

\begin{figure}
\hfill{}\includegraphics[viewport=0bp 0bp 730bp 530bp,clip,width=1\columnwidth]{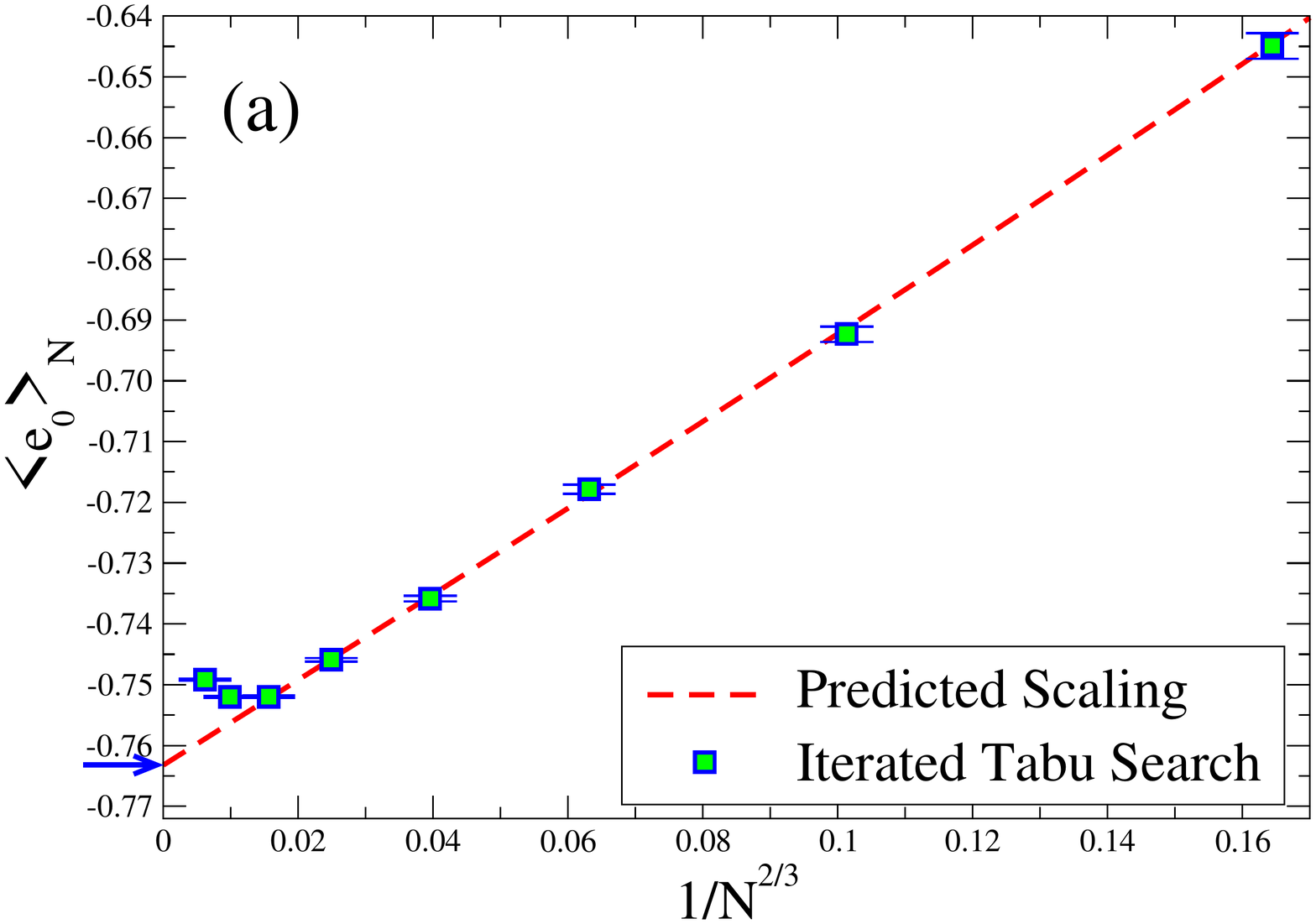}\hfill{}

\hfill{}\includegraphics[viewport=0bp 0bp 730bp 530bp,clip,width=1\columnwidth]{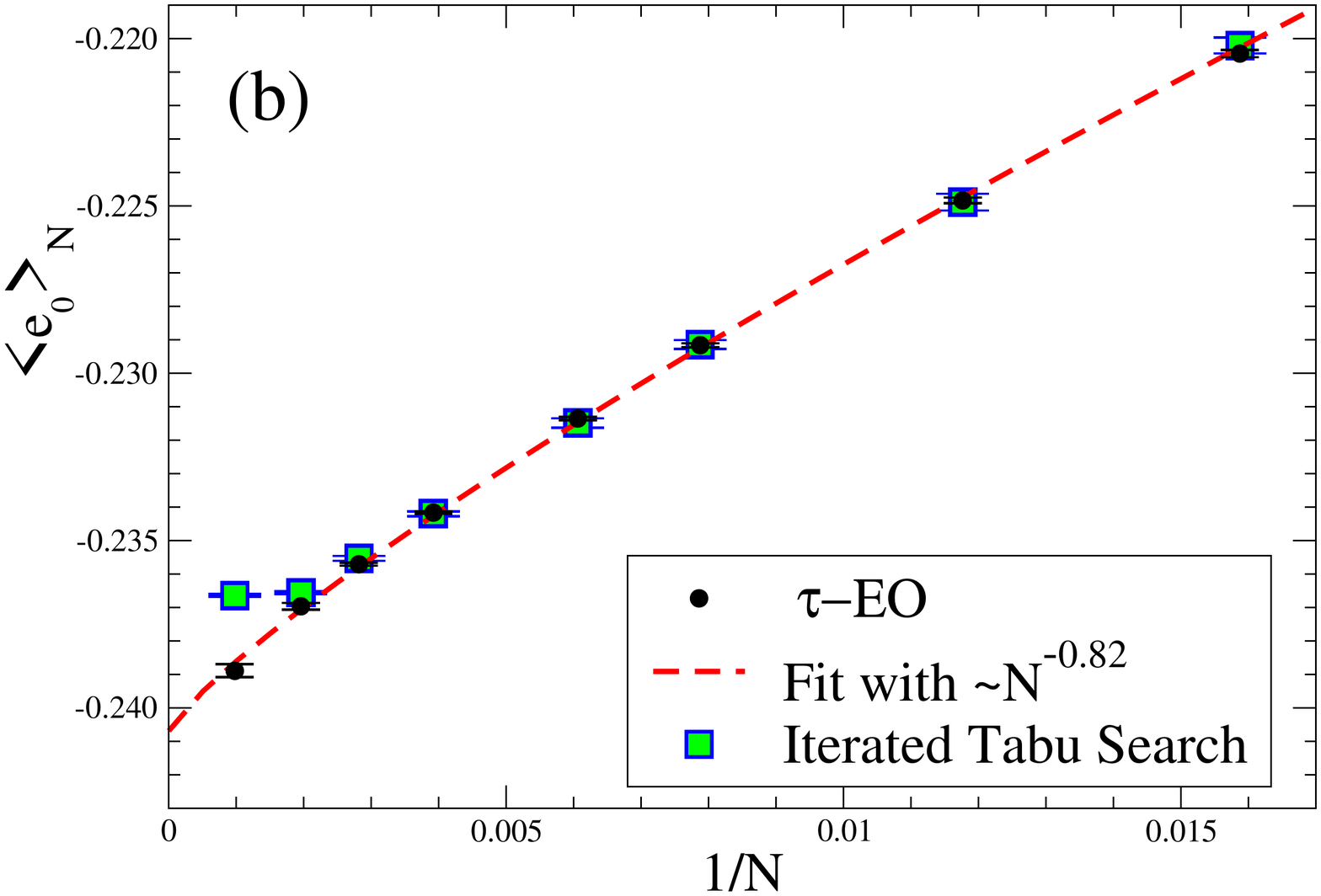}\hfill{}

\vspace{-0.8cm}

\caption{\label{fig:ITSextra}Extrapolation of the average ground-state energy
approximation for the SK, either at full (top) or diluted bond-density
(bottom), as obtained by Iterated Tabu Search~\cite{Palubeckis06},
see data listed in Tab.~\ref{tab:SKextra}. For the full SK on top,
the predicted scaling (red-dashed line) according to Eq.~(\ref{eq:SKextra})
was previously obtained from a fit to an extensive data set obtained
with a different heuristic~\cite{Boettcher10b}. (For example, for
$N\to\infty$ it extrapolates with high accuracy to the exactly known
ground-state energy density of the SK model, $\left\langle e_{0}\right\rangle _{N=\infty}=\text{\textminus}0.763166\ldots$~\cite{Oppermann07,pankov:06},
marked by a blue arrow.) The dilute system on the bottom has not been
studied before, so we have used $\tau-$EO to provide reference data
(black circles), on which the predicted scaling is based via a fit
(red-dashed line) to Eq.~(\ref{eq:SKextra}).}
\end{figure}

This optimization problem of finding ground states of SK has been
tackled previously using genetic algorithms~\cite{Palassini08},
hysteretic optimization~\cite{Pal06b,Goncalves08}, extremal optimization
(EO)~\cite{EOSK,Boettcher10a}, as well as various Metropolis methods~\cite{Grest86,Aspelmeier07}.
In particular, in Refs.~\cite{EOSK,Boettcher10a}, an asymptotic
extrapolation was determined from finite-$N$ data with significant
accuracy for the ensemble-averaged ground state energy,
\begin{equation}
\left\langle e_{0}\right\rangle _{N}=\left\langle e_{0}\right\rangle _{\infty}+\frac{A}{N^{\omega}}\label{eq:SKextra}
\end{equation}
with $\left\langle e_{0}\right\rangle _{\infty}=-0.76323(5)$, $A=0.70(1)$,
and with $\omega=\frac{2}{3}$ conjectured to be exact. It provides
a powerful reference -- alternative to the results obtained from
testbeds -- for the quality of heuristic solvers, as shown in Fig.~\ref{fig:ITSextra}.
There, we plot the results of our simulations where we have averaged
over 1000 instances each for a range of sizes $N$. Those results
are also listed in Tab.~\ref{tab:SKextra}.

As we will compare below with data obtained for QUBO instances in
dilute systems, we supplement this discussion further with a brief
study of SK on a diluted graph. To be comparable with the QUBO instances,
we consider SK in Eq. (\ref{eq:Heq}) with a symmetric bond matrix
$J_{ij}$ whose off-diagonal elements are only to 10\% non-zero (i.e.,
$\pm1$). Again, we have no external fields. Those results are also
listed in Tab.~\ref{tab:SKextra}. These results, also shown in Fig.~\ref{fig:ITSextra},
are practically indistinguishable from those of the full SK. At around
$N\approx500$, ITS exhibits noticeable deviations from the apparent
scaling. Novel to this case is the fact that we can arrive at this
conclusion even though we have no knowledge a-priori about its asymptotic
behavior, which further serves to demonstrate the value of such an
extrapolation in assessing the ability of a heuristic. {[}The fact
that the extrapolation based on our $\tau-$EO data according to Eq.
(\ref{eq:SKextra}) here requires an anomalous exponent of $\omega\approx0.82$
is a novel result in itself and will be studied in more detail elsewhere.{]}

\begin{table}
\caption{\label{tab:SKextra}Average ground state energy obtained for the SK
spin glass, both at full bond-density (left columns) as well as at
a 10\% dilute bond-density (right columns), using the Iterated Tabu
Search heuristic (ITS), as developed for QUBO in Ref.~\cite{Palubeckis06},
by sampling about 1000 instances at each size $N$, and applying default
settings. In the dilute case, we also listed results for $\tau-$EO,
which were generated here just for this comparison. These data points
are also plotted in Fig.~\ref{fig:ITSextra}.}

\hfill{}%
\begin{tabular}{|r|l|r|l|r|l|}
\hline 
\multicolumn{2}{|c|}{Full SK} & \multicolumn{4}{c|}{SK at 10\%}\tabularnewline
\hline 
\multicolumn{2}{|c|}{ITS} & \multicolumn{2}{c|}{ITS} & \multicolumn{2}{c|}{$\tau-$EO}\tabularnewline
\hline 
$N$ & $\left\langle e_{0}\right\rangle _{N}$ & $N$ & $\left\langle e_{0}\right\rangle _{N}$ & $N$ & $\left\langle e_{0}\right\rangle _{N}$\tabularnewline
\hline 
\hline 
15 & -0.644(2) & 63 & -0.2203(3) & 63 & -0.2204(1)\tabularnewline
\hline 
31 & -0.692(1) & 85 & -0.2248(3) & 85 & -0.2248(1)\tabularnewline
\hline 
63 & -0.7178(7) & 127 & -0.2291(2) & 127 & -0.2292(1)\tabularnewline
\hline 
127 & -0.7358(5) & 165 & -0.2314(2) & 165 & -0.2314(1)\tabularnewline
\hline 
255 & -0.7458(3) & 255 & -0.2342(1) & 255 & -0.2342(1)\tabularnewline
\hline 
511 & -0.7519(2) & 355 & -0.2355(1) & 355 & -0.2357(1)\tabularnewline
\hline 
1023 & -0.7520(1) & 511 & -0.2365(1) & 511 & -0.2371(2)\tabularnewline
\hline 
2047 & -0.7491(1) & 1023 & -0.2366(1) & 1023 & -0.2389(3)\tabularnewline
\hline 
\end{tabular}\hfill{}
\end{table}
In either case, for small $N\lesssim512$, the data obtained with
Tabu Search tracks the prediction in Eq.~(\ref{eq:SKextra}) quite
closely, thus demonstrating the consistency with the scaling in Eq.~(\ref{eq:SKextra}).
However, systematic errors become increasingly apparent for larger
system sizes. This raises the following conundrum: Why is a heuristic
like ITS that routinely solves QUBO instances with 10 times as many
variables failing to optimize SK instances beyond 500 variables, considering
the rather similar formulations of both problems? A few immediately
obvious explanations come to mind. For one, the ITS implementation
has been tuned for a certain ensemble, as discussed in Sec.~\ref{sec:RelationSG_QUBO},
while the transformation of SK to QUBO provides a similar but not
identical ensemble. (In fact, ITS specifically employs the strength
of the diagonal $q_{ii}-$weights, which are very distinct in the
SK problem, to initiate its restarts~\cite{Palubeckis06}.) Experts
in Tabu-based heuristics could justifiably argue that with some small
adjustments big improvements can be achieved. In fact, simply increasing
the duration and the number of restarts in ITS leads to a decrease,
albeit slowly, in the systematic error at larger $N$. Yet, the performance
is never quite as impressive as the results obtained by Tabu solvers
for the typical testbed instances of QUBO. We believe that the discrepancy
is the sign of an inherent weakness in the design of the QUBO testbeds.
This is made apparent by showing that heuristics trained on spin glasses
in turn are easily adapted to solve much large samples of QUBO, as
the following discussion suggests. 

\section{Using $\tau-$EO to Solve QUBO Problems\label{sec:Using-EO-to}}

In this section, we proceed to apply heuristic methods developed for
the approximation of spin-glass ground-states to the QUBO problem,
specifically, $\tau-$EO~\cite{Boettcher00,Boettcher01a,Dagstuhl04}.
According to our prior experience, and in contrast to the preceding,
somewhat naive application of the ITS heuristic, we are in a position
to study this implementation in depth and develop a highly tuned heuristic.
On one level, the equivalent spin-glass problem derived from QUBO,
see Sec.~\ref{sec:RelationSG_QUBO}, raises additional challenges
for EO, as the emergence of external fields add new, competing energy
scales to reckon with. However, in the end, the comparison with the
QUBO problem leads us to an understanding and, ultimately, to means
to systematically incorporate these new scales into the search process.
Moreover, an analysis of the solutions obtained for the QUBO problem
as spin glass resolves the conundrum about the size discrepancy in
the solvability of either problem mentioned in the previous section
in physical terms. 

\subsection{Extremal Optimization (EO) Heuristic\label{subsec:Extremal-Optimization-(EO)}}

EO performs a local search~\cite{Hoos04} on an existing configuration
of $N$ variables by changing preferentially those of poor \textit{local}
arrangement. For example, in case of the spin glass model in Eq.~(\ref{eq:Heq}),
but without an external field (i.e., $h_{i}\equiv0$), one usually
sets~\cite{Boettcher01a} $\lambda_{i}=\sigma_{i}\sum_{j}J_{i,j}\sigma_{j}$
to assess the local ``fitness'' of variable $\sigma_{i}$. Then,
$H=-\sum_{i}\lambda_{i}$ represents the overall energy (or cost)
to be minimized. EO simply \emph{ranks} variables, 
\begin{eqnarray}
\lambda_{\Pi(1)}\leq\lambda_{\Pi(2)}\leq\ldots\leq\lambda_{\Pi(N)},\label{rankeq}
\end{eqnarray}
where $\Pi(k)=i$ is the index for the $k^{{\rm th}}$-ranked variable
$\sigma_{i}$. Basic EO~\cite{Boettcher00} always selects the lowest
rank, $k=1$, for an update. Instead, $\tau-$EO selects the $k^{{\rm th}}$-ranked
variable according to a scale-free probability distribution 
\begin{equation}
P(k)\propto k^{-\tau}.\label{eq:taueq}
\end{equation}
The selected variable is updated \emph{unconditionally}, and its fitness
and that of its neighboring variables are reevaluated. This update
is repeated as long as desired, where the unconditional update ensures
significant fluctuations, with sufficient incentive to return to near-optimal
solutions due to selection \emph{against} variables with poor fitness,
for the right choice of $\tau$. Clearly, for finite $\tau$, this
version of EO never ``freezes'' into a single configuration; it
is able to return an extensive list~\cite{Boettcher03a,Boettcher04a}
of the best configurations visited (or simply their cost) ``on the
go'' instead.

For $\tau=0$, the distribution in Eq.~(\ref{eq:taueq}) becomes
flat over the ranks and $\tau-$EO simply becomes a random walk through
configuration space, for which poor search results are to be expected.
Conversely, for $\tau\to\infty$, the process approaches a deterministic
local search, only updating the lowest-ranked variable, $k=1$, and
is likely to get trapped. However, for finite values of $\tau$ the
choice of a \emph{scale-free} distribution for $P(k)$ in Eq.~(\ref{eq:taueq})
ensures that no rank $k$ gets excluded from further evolution, while
maintaining a bias against variables with bad fitness. Fixing $\tau-1\sim1/\ln(N)$
provides a simple, parameter-free strategy, activating avalanches
of adaptation~\cite{eo_jam}.

\begin{figure}
\hspace*{\fill}\includegraphics[viewport=10bp 10bp 750bp 520bp,clip,width=1\columnwidth]{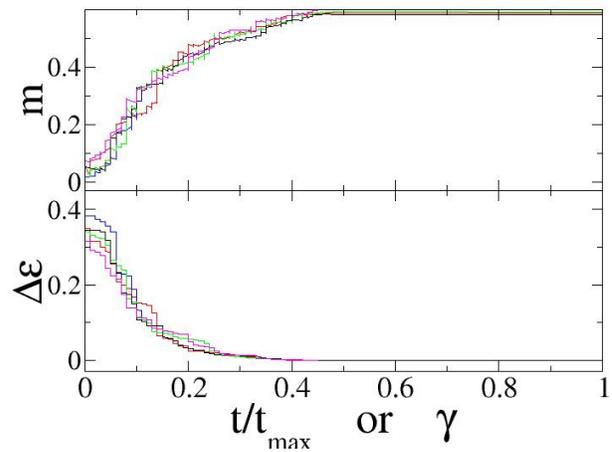}
\hspace*{\fill}

\vspace{-0.7cm}

\caption{\label{fig:timeplot}Plot of the evolution of EO in single runs for
$\tau=1.3$ and $t_{max}=N^{3}/10$ updates, (top) for the magnetization
and (bottom) the relative error $\Delta\epsilon$ with respect to
the best-known value from Ref.~\cite{Glover10} for the instances
of size $N=2500$ in the \emph{bqp} testbed. Starting with a random
assignment of spins at $\Delta\epsilon\approx40\%$, better solutions
are only obtained after the external fields are ramped up somewhat,
according to Eq.~(\ref{eq:lambda_ramped}). But note that the optimal
solution is already found typically when the relative field strength
reaches merely $\gamma\approx50\%$. The fact that the magnetization
of those optimal states reaches $m\approx60\%$, i.e., up to 80\%
of spins simply align with their external field $h_{i}$, indicates
a high degree of redundancy within those instances, see Fig.~\ref{fig:ranklist2500}.
A heuristic merely needs to sort out which 20\% spins have to resist
their external field. }
\end{figure}

\subsection{$\tau-$EO Implementation for QUBO\label{subsec:EO-Implementation-for}}

In light of previous applications to spin glasses, where fitness is
defined via the local field exerted on each spin (see, for example,
Sec.~\ref{subsec:Extremal-Optimization-(EO)}), it would seem straightforward
to simply add the external field $h_{i}$ to the local field to obtain
a definition of fitness as $\lambda_{i}=\sigma_{i}\left(h_{i}+\sum_{j}J_{i,j}\sigma_{j}\right)$,
so that again $H=-\sum_{i}\lambda_{i}$, in accordance with Eq.~(\ref{eq:Heq}).
This canonical approach leads to a problem in which the heuristic
is trying to satisfy two, in principle distinct, scales: that of the
distribution of the bonds $J_{ij}$, and that of the distribution
of the fields $h_{i}$. Since in the QUBO problem both scales derive
from the one distribution of the weights $q_{ij}$, they are correlated
in this case. Yet, in the optimization runs with $\tau-$EO on the
testbed instances~\cite{Boettcher2015}, for example, this definition
of fitnesses $\lambda_{i}$ fails to provide reasonable results. Only
when the external fields were slowly turned on, in those trials, via
a ramp $\gamma$ that is linear in time, 
\begin{equation}
\lambda_{i}=\sigma_{i}\left[\sum_{j=1}^{N}J_{ij}\sigma_{j}+\gamma(t)h_{i}\right],\qquad\gamma(t)=\frac{t}{t_{{\rm max}}},\label{eq:lambda_ramped}
\end{equation}
the best-known results for that testbed were readily reproduced, albeit
at significant overhead in CPU-time. 

In Fig.~\ref{fig:timeplot}, we plot the evolution of the error relative
to that best-known result for each of the 10 instances of the testbed
``bqp2500'', together with the corresponding magnetization. ( ``Magnetization''
here refers to the excess of spins aligned with their external fields
$h_{i}$, whether those are positive or negative. Alternatively, it
may refer to the actual magnetization, $m=\frac{1}{N}\sum_{i=1}^{N}\sigma_{i}$,
due to the excess of spins with $\sigma_{i}=+1$ after applying the
gauge transformation in Sec.~\ref{sub:Gauge-Transformation} that
renders all fields $h_{i}^{\prime}>0$. Both formulations are equivalent!)
At least, two aspects of those results are remarkable. For one, in
each case, the best-found solution is found at least when those fields
are ``turned on'' by 50\%. Secondly, in that best-found solution
there is a high degree of ordering imposed on the instance due to
those external fields. We consider the importance of the latter observation
first.

\begin{figure}
\hspace*{\fill}\includegraphics[viewport=30bp 30bp 750bp 540bp,clip,width=1\columnwidth]{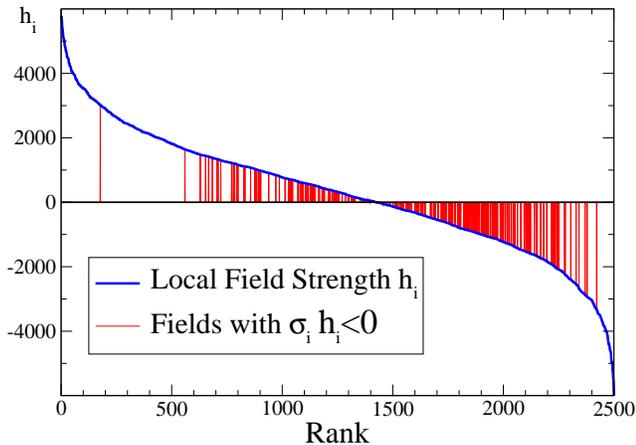}
\hspace*{\fill}

\vspace{-0.5cm}

\caption{\label{fig:ranklist2500}Analysis of the magnetization of the best-known
solution to one of the \emph{bqp2500}-instances from the QUBO testbed.
Here, the rank-ordered list of $N=2500$ local fields $h_{i}$ defined
in Eq.~(\ref{eq:Jh_QUBO}), corresponding to the row/column-sum of
the weights $q_{ij}$, are plotted (blue line). Marked (by red vertical
lines) are fields where the associated spin $\sigma_{i}$ in the configuration
with the lowest energy is \emph{not} aligned with $h_{i}$, i.e.,
when $\sigma_{i}h_{i}<0$. As the results for the magnetization in
Fig.~\ref{fig:timeplot} suggest, only a small fraction ($\approx20\%$)
of variables do not align, in particular, most of those associated
with the (absolute) highest fields (and, thus, largest contributions
to the energy) are aligned (unmarked) with high probability. }
\end{figure}
\begin{figure}
\hspace*{\fill}\includegraphics[viewport=30bp 30bp 750bp 540bp,clip,width=1\columnwidth]{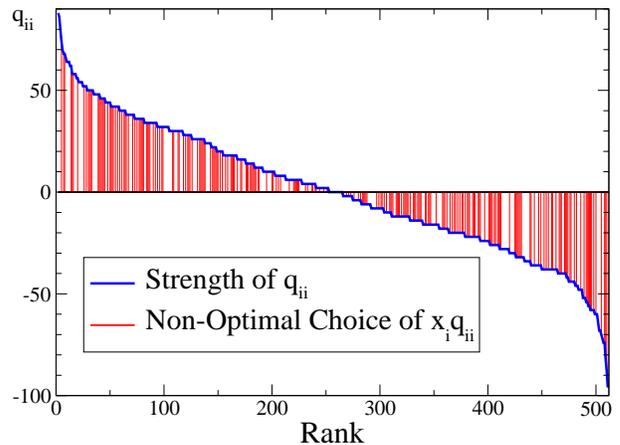}
\hspace*{\fill}

\vspace{-0.5cm}

\caption{\label{fig:SKranklist500}Analysis of the coercion of the diagonal
element $q_{ii}$ in the QUBO matrix obtained from an SK spin-glass
instance of size $N=511$ on the Boolean variables $x_{i}$ in the
best-known solution. As in Fig. \ref{fig:ranklist2500}, the $q_{ii}$,
obtained via Eq. (\ref{eq:qij_SK}), are plotted (in blue) in rank-order.
Marked (by red vertical lines) are fields where the associated variable
$x_{i}$ in the configuration with the lowest cost is \emph{not} aligned
to render $x_{i}q_{ii}$ optimal, i.e., when $x_{i}=0$ although $q_{ii}>0$
or $x_{i}=1$ while $q_{ii}<0$ . Unlike the local fields $h_{i}$
in Fig. \ref{fig:ranklist2500}, the strength of $q_{ii}$ here provides
no distinguishable coersive force of the $x_{i}$.}
\end{figure}

\subsubsection{Magnetization of QUBO Instances:\label{subsec:Magnatization-of-QUBO}}

\begin{figure}
\hfill{}\includegraphics[viewport=0bp 0bp 730bp 530bp,clip,width=1\columnwidth]{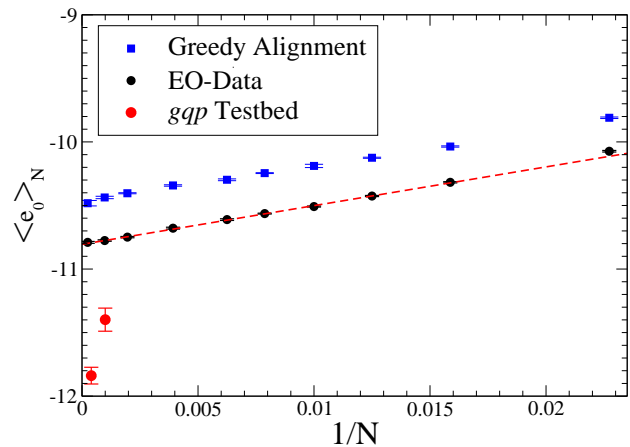}\hfill{}

\vspace{-0.5cm}

\caption{\label{fig:QUBOextra}Extrapolation of the average optimal cost approximation
for the QUBO problem as obtained by $\tau-$EO. All data displayed
here can also be found in Tab.~\ref{tab:QUBOdata}. We can fit the
EO-data (black circles) for sufficiently small $N$ (which is more
likely exact!), but $N\gtrsim50$ to be asymptotic, to obtain a scaling
prediction (red-dashed line) for all large $N$. A deviation from
that scaling, which would signal the onset of systematic errors in
the heuristic (as seen in Fig.~\ref{fig:ITSextra}), is not apparent
here for the data up to $N=4095$. Shown are also the corresponding
data for the greedy alignment algorithm mentioned in the text (blue
squares), which remains systematically 3\% above the optimal results
for all $N$, another indication that the EO-data maintains its systematic
accuracy. Strikingly, the averages for the best-found solutions for
both testbeds, gqp1000 and gqp2500, are uncharacteristically low and
far from their expected ensemble average (red-dashed line), but so
are their greedy approximations (not shown here, but see Tab.~\ref{tab:QUBOdata}),
which are again 3\% higher, yet, much below the ensemble. It seems
likely that those \emph{gap testbed} instances, originating in 1996
from Ref.~\cite{Beasley96}, were generated with a poor random number
generator.}
\end{figure}

Representing each testbed instance of QUBO as a spin glass, following
Sec. \ref{subsec:QUBO-Problem-as}, we actually find that the magnetization
reaches $\approx60\%$, i.e., the alignment of the variables $\sigma_{i}$
with their external fields is to 80\% a predictor of the optimal arrangement
within the lowest-energy solution. Thus, irrespective of the mutual
constraints spins impose on each other through the bonds $J_{ij}$,
in many cases those constraints are simply overwritten by the torque
exerted by the external fields $h_{i}$ alone. Clearly, a larger local
field imposes a larger torque that is more likely coercive than a
smaller one, as Fig.~\ref{fig:ranklist2500} illustrates. In fact,
we find that a simple $O(N)$ ``greedy alignment'' algorithm that
aligns spins sequentially, selected based on having the largest remaining
local field (consisting of the torque exerted by the external field
\emph{and} those of any previously assigned spins), typically reaches
a cost that is within 3\% of the best-known solutions (see Fig.~\ref{fig:QUBOextra}).
Still, it likely remains an NP-hard task to sort out which 20\% of
the fields are to be disobeyed, although for each $N$ this is a problem
of much reduced complexity compared to the corresponding SK ground
state problem with all $h_{i}\equiv0$, hence, explaining the discrepancy
in ``hardness'' between QUBO and SK.

It is instructive to consider \footnote{We thank the referee for insisting on this consideration.}
the -- seemingly -- equivalent representation of an SK-instance
(without external field) as a QUBO problem. As mentioned in Sec. \ref{subsec:Spin-Glass-as},
such a conversion produces a linear term in the QUBO cost-function
of similar appearance to the magnetic field above. In particular,
its strength, given by $q_{ii}$ in Eq. (\ref{eq:qij_SK}), appears
to be as extensive as we found for the $h_{i}$. However, the coupling
of each variable $x_{i}$ to $q_{ii}$ is somewhat arbitrary and,
hence, fails to coerce $x_{i}$ in a significant manner, as is demonstrated
in Fig. \ref{fig:SKranklist500}. To show this, we employ the freedom
to gauge the SK-instance in question, following Sec. \ref{sub:Gauge-Transformation},
leaving open the choice for the entire set of gauge-parameters $\left\{ \xi_{i}\right\} $.
In the absence of an external field, Eq. (\ref{eq:qij_SK}) then yields:
\begin{equation}
q_{ii}=-\sum_{l=1}^{N}J_{il}\xi_{i}\xi_{l}.\label{eq:gaugedQii}
\end{equation}
Alas, different choices of $\xi_{i}=\pm1$ create quite arbitrary
linear couplings $q_{ii}$ for each individual $x_{i}$ (although
overall the hardness of the problem is not affected)! In contrast,
no such invariance exists for QUBO and, thus, the linear terms $h_{i}$
emerging in its conversion into a spin-glass problem are unique and
render the coercive force on their coupled variable $\sigma_{i}$
consequential, see Fig. \ref{fig:ranklist2500}.

\subsubsection{Optimization of the EO-Implementation:\label{subsec:Optimization-of-the}}

\begin{figure}
\hspace*{\fill}\includegraphics[viewport=20bp 20bp 750bp 540bp,clip,width=1\columnwidth]{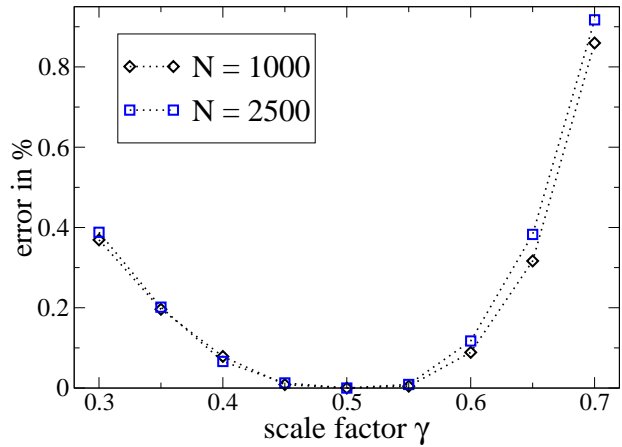}
\hspace*{\fill}

\vspace{-0.5cm}

\caption{\label{fig:dep_gamma}Study of the $\gamma-$dependence of the error
$\epsilon$ produced by $\tau-$EO relative to the best-known results,
averaged over the two QUBO \emph{gqp testbeds} with $N=1000$ and
$N=2500$, when using the fitnesses $\lambda_{i}=\sigma_{i}\left[\sum_{j=1}^{N}J_{ij}\sigma_{j}+\gamma h_{i}\right]$
with fixed $\gamma$ during a run, as generalization of Eq.~(\ref{eq:lambdaSK}).
For $\tau-$EO, as described in Sec.~\ref{subsec:Extremal-Optimization-(EO)},
we set $\tau=1.3$ and $t_{max}=N^{3}/100$. The results clearly indicate
$\gamma=50\%$ as an especially useful case, as implemented in Eq.~(\ref{eq:lambdaSK}). }
\end{figure}

We now return to the earlier observation about $\tau-$EO saturating
the best-known results in the testbed when the ramped fields in Eq.~(\ref{eq:lambda_ramped})
reach 50\% with striking consistency. As it turns out, this observation
pins down an arbitrary choice in the design of EO that allows us to
implement a more efficient version of $\tau-$EO. This choice in the
definition of fitness attributed to individual variables has been
discussed previously in Ref.~\cite{Dagstuhl04}. It is here where
the interpretation of spin glasses as a QUBO problem has its most
significant impact. Unlike for a spin glass, where the combined local
field offers itself as the canonical fitness for each spin, in QUBO
we would naturally construct a fitness as follows instead: By assigning
a variable $x_{i}$, its instantaneous contribution to the cost of
$r_{i}=\sum_{j=1}^{N}q_{ij}x_{j}$ is either suppressed ($x_{i}=0$)
or added ($x_{i}=1$), hence, the fitness $\lambda_{i}$ should be
$r_{i}$ if $x_{i}=1$ or $-r_{i}$ if $x_{i}=0$, penalizing the
un-actualized potential when $r_{i}>0$ but $x_{i}=0$. Thus, for
QUBO the apparent choice for fitness can be summarized as 
\begin{equation}
\lambda_{i}=\sigma_{i}\sum_{j=1}^{N}q_{ij}x_{j}.\label{eq:lambdaQUBO}
\end{equation}
Note that in this case, $\sum_{i}\lambda_{i}$ itself does not add
up to the actual cost of an instance, $E$ or $H$, which is not a
necessity, as is discussed in Ref.~\cite{Dagstuhl04}. Amazingly,
using the definitions in Sec.~\ref{sec:RelationSG_QUBO}, for the
spin glass this translates into
\begin{equation}
\lambda_{i}=\sigma_{i}\left[\sum_{j=1}^{N}J_{ij}\sigma_{j}+\frac{1}{2}h_{i}\right],\label{eq:lambdaSK}
\end{equation}
i.e., favoring a fixed value of $\gamma=50\%$. This result is indeed
borne out with a more systematic study at various fixed values of
$\gamma$, as shown in Fig.~\ref{fig:dep_gamma}. Accordingly, we
will use this more effective version of $\tau-$EO in the following,
with $\tau=1.3$ and $t_{max}=N^{3}/100$, and fitnesses as defined
in Eq.~(\ref{eq:lambdaSK}), to study the QUBO problem as a spin
glass. As such, $\tau-$EO has a complexity of $O\left(N^{3}\ln N\right)$,
where the logarithmic dependence is due to dynamic sorting of fitnesses,
as introduced in Ref.~\cite{EOSK}.

\begin{table}
\caption{\label{tab:QUBOdata}Results from applying $\tau-$EO to the QUBO
ensemble defined in Sec.~\ref{sec:RelationSG_QUBO}, with $q_{ij}$
drawn randomly from a flat distribution over the integers on $\left[-100\ldots+100\right]$
at 10\% filling. Listed are the system sizes $N$ considered, the
number of instances $I$ simulated from the ensemble, the measured
ground-state energy density $\left\langle e_{0}\right\rangle _{N}=H/N^{\frac{3}{2}}$
according to Eq.~(\ref{eq:Heq}), and the corresponding approximation
obtained with the greedy alignment algorithm. Note that the result
for $N=1000$ and $N=2500$ specifically refer only to the \emph{gap
testbeds} (underlined). This data is plotted as an extrapolation plot
in Fig.~\ref{fig:QUBOextra}.}

\hfill{}%
\begin{tabular}{|r|r|l|l|}
\hline 
$N$ & $I$ & $\left\langle e_{0}\right\rangle _{N}$ & Greedy\tabularnewline
\hline 
\hline 
31 & $10^{5}$ & -9.67(1) & -9.46(1)\tabularnewline
\hline 
44 & $10^{5}$ & -10.074(5) & -9.81(1)\tabularnewline
\hline 
63 & $10^{5}$ & -10.318(5) & -10.036(5)\tabularnewline
\hline 
80 & $10^{5}$ & -10.426(4) & -10.125(4)\tabularnewline
\hline 
100 & $10^{5}$ & -10.501(4) & -10.190(5)\tabularnewline
\hline 
127 & $10^{5}$ & -10.564(3) & -10.245(3)\tabularnewline
\hline 
160 & $2\,10^{4}$ & -10.611(7) & -10.298(7)\tabularnewline
\hline 
255 & $10^{4}$ & -10.68(1) & -10.34(1)\tabularnewline
\hline 
511 & $10^{4}$ & -10.750(5) & -10.404(4)\tabularnewline
\hline 
\uuline{1000} & \uuline{10} & \uuline{-11.4(1)} & \uuline{-11.1(1)}\tabularnewline
\hline 
1023 & $4\,10^{3}$ & -10.776(6) & -10.44(1)\tabularnewline
\hline 
\uuline{2500} & \uuline{10} & \uuline{-11.84(7)} & \uuline{-11.45(7)}\tabularnewline
\hline 
4095 & 600 & -10.79(1) & -10.48(2)\tabularnewline
\hline 
\end{tabular}\hfill{}
\end{table}

\subsection{Ensemble Results for the QUBO Problem\label{subsec:Ensemble-Results-for}}

Based on the implementation of $\tau-$EO described in the previous
section, we have run extensive simulations for the QUBO problem, similar
to those we have employed previously for SK~\cite{EOSK,Boettcher10b}.
And in analogy with those, we propose here to evaluate the capabilities
of the implementation using an extrapolation plot of the ensemble
results, as also shown in Fig.~\ref{fig:ITSextra}. The results validate
our expectation that QUBO problems in this ensemble can be solved
to much larger sizes than the corresponding SK spin glass.

In Tab.~\ref{tab:QUBOdata}, we summarize the results of the simulations
for the range of instance sizes from $N=31,\ldots,4095$. For each
size, we have selected a sufficiently large number of instances from
the ensemble to be able to keep the statistical errors small and relatively
comparable in magnitude. From SK, it is well-known that, if the matrix
elements are drawn from a distribution of fixed width, scale-invariant
(intensive) costs are obtained when $H$ is rescaled by a factor of
$N^{\frac{3}{2}}$~\cite{MPV}, thus, we define $\left\langle e_{0}\right\rangle _{N}=H/N^{\frac{3}{2}}$,
in accordance with Eq.~(\ref{eq:SKextra}). Listed are also the corresponding
results for the described $O(N)$ Greedy Alignment algorithm, which
turn out to be consistently 3\% above the best EO predictions, another
sign of their systematic quality. 

\begin{figure}
\hfill{}\includegraphics[viewport=0bp 0bp 730bp 530bp,clip,width=1\columnwidth]{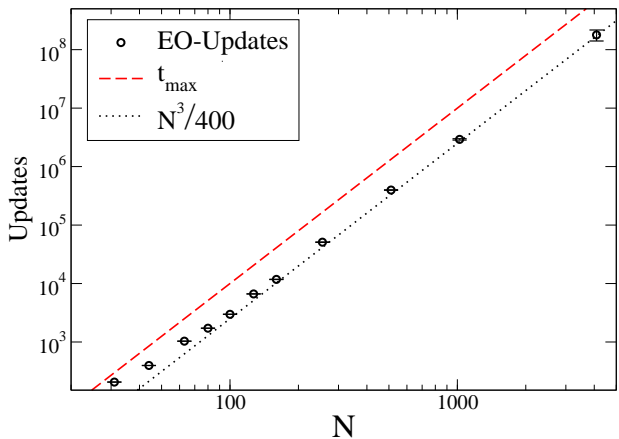}\hfill{}

\caption{\label{fig:CPUcostEO}Measure of the computational cost for the $\tau-$EO
implementation in terms of the average number of update steps needed
to first encounter the best-found solutions listed in Tab.~\ref{tab:QUBOdata}
as a function of instance size $N$. The cubic line (dotted) of $N^{3}/400$
is merely included to guide the eye. The approach of the measured
updates suggest that EO typically finds its best solution within a
quarter of the allotted number of updates, $t_{{\rm max}}=N^{3}/100$
(red-dashed line). Since the actual computational complexity for an
NP-hard problem such as QUBO is expected to rise exponentially in
size, we would expect EO to eventually exhibit systematic errors.
However, up to this size, there is no sign of upward pressure on the
total runtime.}
\end{figure}

This data is also plotted in Fig.~\ref{fig:QUBOextra}, in extrapolated
form, which should yield an asymptotically linear graph, according
to Eq.~(\ref{eq:SKextra}), if we choose $N^{-\omega}$ with the
correct value of $\omega$ as our $x-$axis. Such a linear extrapolation
is achieved here for $\omega=1$, suggesting that finite-size corrections
in QUBO diminish much faster than for SK, where corrections are conjectured
to decay only as $N^{-\frac{2}{3}}$, i.e., $\omega=\frac{2}{3}$~\cite{EOSK,Billoire06,Aspelmeier07},
as shown in Fig.~\ref{fig:ITSextra}. Weaker corrections provide
more evidence for the relative simplicity of approximating QUBO. As
for the SK data in Refs.~\cite{EOSK,Boettcher10b}, this data is
also readily fitted asymptotically (for $N$ small enough that there
a few systematic errors but large enough, here $N>44$, to ignore
finite-size corrections) with the linear form provided by Eq.~(\ref{eq:SKextra}).
Note that the specific values obtained for this fit, $\left\langle e_{0}\right\rangle _{\infty}=-10.8(1)$
and $A=30(1)$, are not of any significance by themselves. All we
care about is a deviation from that line for large $N$ as a likely
sign of a systematic breakdown in the heuristic we care to assess.
Up to the sizes accessible with this implementation within reasonable
CPU time, EO does not show any significant systematic error, as discussed
in Fig. \ref{fig:CPUcostEO}.

As a curious side-note, we observe that the 10 instances from the
\emph{gap testbeds} of sizes $N=1000$ and $N=2500$ (also listed
in Tab.~\ref{tab:QUBOdata} and plotted as red dots in Fig.~\ref{fig:QUBOextra})
apparently are highly atypical for the ensemble they were supposedly
drawn from, with much lower average costs. This does \emph{not} signal
a shortcoming of EO, as all averages were obtained uniformly with
the same implementation, and the greedy results are equally untypical
but remain 3\% above the best-found costs. We can only speculate about
the origin of this effect, but it seems likely that a poor random
number generator was used to make the testbed. 

\section{Conclusions\label{sec:Conclusions}}

In our discussion, we have analyzed the relation between the SK spin
glass ground-state problem and the classical NP-hard combinatorial
problem of QUBO. We have argued that a widely used form of the QUBO
problem, with weights drawn from a symmetric distribution of finite
width, leads to rather simple testbeds with a high degree of redundancy.
Those instances correspond to a spin glass problem where a large fraction
of spins are independently determined by a large biasing field. As
Eq. (\ref{eq:Jh_QUBO}) shows, those biasing fields can only be avoided
when in the QUBO problem the sum of the weights in each row (or column)
vanishes, or at least grows less that $O(\sqrt{N})$ for increasing
problem size $N$. We will explore more systematic approaches to generate
matrices $Q$ that have random entries but are constraint to vanishing
row-sums elsewhere. Since such problems have recently been used to
assess the quality of dedicated quantum annealers~\cite{McGeoch13}
such as D-Wave, which claims advantages due to quantum effects, a
careful analysis of actual hardness of classical problems is timely.
In terms of the physical description of the QUBO problem as a Ising
spin glass, we find that a large fraction of variables in those instances
are trivially coerced by large external fields. The impact of this
redundancy is illustrated first by applying a standard QUBO solver
that provides good results for large QUBO instances but in turn fails
for much smaller and seemingly similar SK instances. We then proposed
an implementation of $\tau-$EO, previously well-trained on the SK
problem, and show that it can solve comparatively much larger instances
of the QUBO problem. Along the way, we have shown that a systematic,
ensemble-based study to test the capabilities of heuristics via an
extrapolation plot provides a self-contained and quite stringent measure
of their performance for large $N$, superior to any ad-hoc assembly
of testbeds.

In the future, we will explore whether the definition of fitness used
in Eq.~(\ref{eq:lambdaSK}) for spin glasses in an external field,
which our calculations show to remain valid when the external field
is varied independently (unlike for the SK obtained from QUBO here),
will allow to apply $\tau-$EO also to interesting ground state problems
of spin glasses in such fields. A number of questions about the low-temperature
glassy state of spin glasses are connected with its stability under
coercion with external fields~\cite{Almeida78,Joerg08,Larson13,Zintchenko14}.

\section*{Acknowledgements}

The author likes to thank Prof.~Gintaras Palubeckis for his generous
permission to use his implementation of Iterated Tabu Search (ITS)
from his webpage (https://www.personalas.ktu.lt/\textasciitilde ginpalu/)
at Kaunas University of Technology. The author also thanks Dr.~Matthias
Troyer for suggesting the application of EO to the QUBO problem.

\bibliographystyle{apsrev4-1}
\bibliography{/Users/sboettc/Boettcher}

\begin{thebibliography}{46}%
\makeatletter
\providecommand \@ifxundefined [1]{%
 \@ifx{#1\undefined}
}%
\providecommand \@ifnum [1]{%
 \ifnum #1\expandafter \@firstoftwo
 \else \expandafter \@secondoftwo
 \fi
}%
\providecommand \@ifx [1]{%
 \ifx #1\expandafter \@firstoftwo
 \else \expandafter \@secondoftwo
 \fi
}%
\providecommand \natexlab [1]{#1}%
\providecommand \enquote  [1]{``#1''}%
\providecommand \bibnamefont  [1]{#1}%
\providecommand \bibfnamefont [1]{#1}%
\providecommand \citenamefont [1]{#1}%
\providecommand \href@noop [0]{\@secondoftwo}%
\providecommand \href [0]{\begingroup \@sanitize@url \@href}%
\providecommand \@href[1]{\@@startlink{#1}\@@href}%
\providecommand \@@href[1]{\endgroup#1\@@endlink}%
\providecommand \@sanitize@url [0]{\catcode `\\12\catcode `\$12\catcode
  `\&12\catcode `\#12\catcode `\^12\catcode `\_12\catcode `\%12\relax}%
\providecommand \@@startlink[1]{}%
\providecommand \@@endlink[0]{}%
\providecommand \url  [0]{\begingroup\@sanitize@url \@url }%
\providecommand \@url [1]{\endgroup\@href {#1}{\urlprefix }}%
\providecommand \urlprefix  [0]{URL }%
\providecommand \Eprint [0]{\href }%
\providecommand \doibase [0]{http://dx.doi.org/}%
\providecommand \selectlanguage [0]{\@gobble}%
\providecommand \bibinfo  [0]{\@secondoftwo}%
\providecommand \bibfield  [0]{\@secondoftwo}%
\providecommand \translation [1]{[#1]}%
\providecommand \BibitemOpen [0]{}%
\providecommand \bibitemStop [0]{}%
\providecommand \bibitemNoStop [0]{.\EOS\space}%
\providecommand \EOS [0]{\spacefactor3000\relax}%
\providecommand \BibitemShut  [1]{\csname bibitem#1\endcsname}%
\let\auto@bib@innerbib\@empty
\bibitem [{\citenamefont {L\"u}\ \emph {et~al.}(2010)\citenamefont {L\"u},
  \citenamefont {Glover},\ and\ \citenamefont {Hao}}]{Lu10}%
  \BibitemOpen
  \bibfield  {author} {\bibinfo {author} {\bibfnamefont {Z.}~\bibnamefont
  {L\"u}}, \bibinfo {author} {\bibfnamefont {F.~S.}\ \bibnamefont {Glover}}, \
  and\ \bibinfo {author} {\bibfnamefont {J.-K.}\ \bibnamefont {Hao}},\
  }\href@noop {} {\bibfield  {journal} {\bibinfo  {journal} {European Journal
  of Operational Research}\ }\textbf {\bibinfo {volume} {207}},\ \bibinfo
  {pages} {1254} (\bibinfo {year} {2010})}\BibitemShut {NoStop}%
\bibitem [{\citenamefont {McGeoch}\ and\ \citenamefont
  {Wang}(2013)}]{McGeoch13}%
  \BibitemOpen
  \bibfield  {author} {\bibinfo {author} {\bibfnamefont {C.~C.}\ \bibnamefont
  {McGeoch}}\ and\ \bibinfo {author} {\bibfnamefont {C.}~\bibnamefont {Wang}},\
  }in\ \href {\doibase 10.1145/2482767.2482797} {\emph {\bibinfo {booktitle}
  {Proceedings of the ACM International Conference on Computing Frontiers}}},\
  \bibinfo {series and number} {CF '13}\ (\bibinfo  {publisher} {ACM},\
  \bibinfo {address} {New York, NY, USA},\ \bibinfo {year} {2013})\ pp.\
  \bibinfo {pages} {23:1--23:11}\BibitemShut {NoStop}%
\bibitem [{\citenamefont {Aramon}\ \emph {et~al.}(2019)\citenamefont {Aramon},
  \citenamefont {Rosenberg}, \citenamefont {Valiante}, \citenamefont
  {Miyazawa}, \citenamefont {Tamura},\ and\ \citenamefont
  {Katzgraber}}]{Aramon2019}%
  \BibitemOpen
  \bibfield  {author} {\bibinfo {author} {\bibfnamefont {M.}~\bibnamefont
  {Aramon}}, \bibinfo {author} {\bibfnamefont {G.}~\bibnamefont {Rosenberg}},
  \bibinfo {author} {\bibfnamefont {E.}~\bibnamefont {Valiante}}, \bibinfo
  {author} {\bibfnamefont {T.}~\bibnamefont {Miyazawa}}, \bibinfo {author}
  {\bibfnamefont {H.}~\bibnamefont {Tamura}}, \ and\ \bibinfo {author}
  {\bibfnamefont {H.~G.}\ \bibnamefont {Katzgraber}},\ }\href {\doibase
  10.3389/fphy.2019.00048} {\bibfield  {journal} {\bibinfo  {journal}
  {Frontiers in Physics}\ }\textbf {\bibinfo {volume} {7}} (\bibinfo {year}
  {2019}),\ 10.3389/fphy.2019.00048}\BibitemShut {NoStop}%
\bibitem [{\citenamefont {Wang}\ \emph {et~al.}(2013)\citenamefont {Wang},
  \citenamefont {L\"u}, \citenamefont {Glover},\ and\ \citenamefont
  {Hao}}]{Wang13}%
  \BibitemOpen
  \bibfield  {author} {\bibinfo {author} {\bibfnamefont {Y.}~\bibnamefont
  {Wang}}, \bibinfo {author} {\bibfnamefont {Z.}~\bibnamefont {L\"u}}, \bibinfo
  {author} {\bibfnamefont {F.}~\bibnamefont {Glover}}, \ and\ \bibinfo {author}
  {\bibfnamefont {J.-K.}\ \bibnamefont {Hao}},\ }\href {\doibase
  10.1016/j.cor.2011.12.006} {\bibfield  {journal} {\bibinfo  {journal}
  {Computers {\&} Operations Research}\ }\textbf {\bibinfo {volume} {40}},\
  \bibinfo {pages} {3100} (\bibinfo {year} {2013})}\BibitemShut {NoStop}%
\bibitem [{\citenamefont {Glover}\ \emph {et~al.}(2010)\citenamefont {Glover},
  \citenamefont {L\"u},\ and\ \citenamefont {Hao}}]{Glover10}%
  \BibitemOpen
  \bibfield  {author} {\bibinfo {author} {\bibfnamefont {F.~S.}\ \bibnamefont
  {Glover}}, \bibinfo {author} {\bibfnamefont {Z.}~\bibnamefont {L\"u}}, \ and\
  \bibinfo {author} {\bibfnamefont {J.-K.}\ \bibnamefont {Hao}},\ }\href@noop
  {} {\bibfield  {journal} {\bibinfo  {journal} {4OR - Q J Oper Res}\ }\textbf
  {\bibinfo {volume} {8}},\ \bibinfo {pages} {239} (\bibinfo {year}
  {2010})}\BibitemShut {NoStop}%
\bibitem [{\citenamefont {Boros}\ \emph {et~al.}(2007)\citenamefont {Boros},
  \citenamefont {Hammer},\ and\ \citenamefont {Tavares}}]{Boros07}%
  \BibitemOpen
  \bibfield  {author} {\bibinfo {author} {\bibfnamefont {E.}~\bibnamefont
  {Boros}}, \bibinfo {author} {\bibfnamefont {P.~L.}\ \bibnamefont {Hammer}}, \
  and\ \bibinfo {author} {\bibfnamefont {G.}~\bibnamefont {Tavares}},\
  }\href@noop {} {\bibfield  {journal} {\bibinfo  {journal} {J. Heuristics}\
  }\textbf {\bibinfo {volume} {13}},\ \bibinfo {pages} {99} (\bibinfo {year}
  {2007})}\BibitemShut {NoStop}%
\bibitem [{\citenamefont {Palubeckis}(2006)}]{Palubeckis06}%
  \BibitemOpen
  \bibfield  {author} {\bibinfo {author} {\bibfnamefont {G.}~\bibnamefont
  {Palubeckis}},\ }\href {http://dl.acm.org/citation.cfm?id=1413841.1413850}
  {\bibfield  {journal} {\bibinfo  {journal} {Informatica}\ }\textbf {\bibinfo
  {volume} {17}},\ \bibinfo {pages} {279} (\bibinfo {year} {2006})}\BibitemShut
  {NoStop}%
\bibitem [{\citenamefont {Kochenberger}\ \emph {et~al.}(2014)\citenamefont
  {Kochenberger}, \citenamefont {Hao}, \citenamefont {Glover}, \citenamefont
  {Lewis}, \citenamefont {L\"u}, \citenamefont {Wang},\ and\ \citenamefont
  {Wang}}]{Kochenberger2014}%
  \BibitemOpen
  \bibfield  {author} {\bibinfo {author} {\bibfnamefont {G.}~\bibnamefont
  {Kochenberger}}, \bibinfo {author} {\bibfnamefont {J.-K.}\ \bibnamefont
  {Hao}}, \bibinfo {author} {\bibfnamefont {F.}~\bibnamefont {Glover}},
  \bibinfo {author} {\bibfnamefont {M.}~\bibnamefont {Lewis}}, \bibinfo
  {author} {\bibfnamefont {Z.}~\bibnamefont {L\"u}}, \bibinfo {author}
  {\bibfnamefont {H.}~\bibnamefont {Wang}}, \ and\ \bibinfo {author}
  {\bibfnamefont {Y.}~\bibnamefont {Wang}},\ }\href {\doibase
  10.1007/s10878-014-9734-0} {\bibfield  {journal} {\bibinfo  {journal}
  {Journal of Combinatorial Optimization}\ }\textbf {\bibinfo {volume} {28}},\
  \bibinfo {pages} {58} (\bibinfo {year} {2014})}\BibitemShut {NoStop}%
\bibitem [{\citenamefont {Barahona}\ \emph {et~al.}(1988)\citenamefont
  {Barahona}, \citenamefont {Gr{\"o}tschel}, \citenamefont {J{\"u}nger},\ and\
  \citenamefont {Reinelt}}]{Barahona88}%
  \BibitemOpen
  \bibfield  {author} {\bibinfo {author} {\bibfnamefont {F.}~\bibnamefont
  {Barahona}}, \bibinfo {author} {\bibfnamefont {M.}~\bibnamefont
  {Gr{\"o}tschel}}, \bibinfo {author} {\bibfnamefont {M.}~\bibnamefont
  {J{\"u}nger}}, \ and\ \bibinfo {author} {\bibfnamefont {G.}~\bibnamefont
  {Reinelt}},\ }\href@noop {} {\bibfield  {journal} {\bibinfo  {journal} {Oper.
  Res.}\ }\textbf {\bibinfo {volume} {36}},\ \bibinfo {pages} {493} (\bibinfo
  {year} {1988})}\BibitemShut {NoStop}%
\bibitem [{\citenamefont {Hartmann}\ and\ \citenamefont
  {Rieger}(2004)}]{Dagstuhl04}%
  \BibitemOpen
  \bibinfo {editor} {\bibfnamefont {A.}~\bibnamefont {Hartmann}}\ and\ \bibinfo
  {editor} {\bibfnamefont {H.}~\bibnamefont {Rieger}},\ eds.,\ \href@noop {}
  {\emph {\bibinfo {title} {New Optimization Algorithms in Physics}}}\
  (\bibinfo  {publisher} {Wiley-VCH},\ \bibinfo {address} {Berlin},\ \bibinfo
  {year} {2004})\BibitemShut {NoStop}%
\bibitem [{\citenamefont {Hoos}\ and\ \citenamefont
  {St{\"u}tzle}(2004)}]{Hoos04}%
  \BibitemOpen
  \bibfield  {author} {\bibinfo {author} {\bibfnamefont {H.~H.}\ \bibnamefont
  {Hoos}}\ and\ \bibinfo {author} {\bibfnamefont {T.}~\bibnamefont
  {St{\"u}tzle}},\ }\href@noop {} {\emph {\bibinfo {title} {Stochastic Local
  Search: Foundations and Applications}}}\ (\bibinfo  {publisher} {Morgan
  Kaufmann},\ \bibinfo {address} {San Francisco},\ \bibinfo {year}
  {2004})\BibitemShut {NoStop}%
\bibitem [{\citenamefont {Osman}\ and\ \citenamefont {Kelly}(1996)}]{Osman96}%
  \BibitemOpen
  \bibinfo {editor} {\bibfnamefont {I.~H.}\ \bibnamefont {Osman}}\ and\
  \bibinfo {editor} {\bibfnamefont {J.~P.}\ \bibnamefont {Kelly}},\ eds.,\
  \href@noop {} {\emph {\bibinfo {title} {Meta-Heuristics: Theory and
  Application}}}\ (\bibinfo  {publisher} {Kluwer},\ \bibinfo {address}
  {Boston},\ \bibinfo {year} {1996})\BibitemShut {NoStop}%
\bibitem [{\citenamefont {Martello}\ \emph {et~al.}(1999)\citenamefont
  {Martello}, \citenamefont {Osman}, \citenamefont {Roucairol},\ and\
  \citenamefont {Voss}}]{Voss99}%
  \BibitemOpen
  \bibinfo {editor} {\bibfnamefont {S.}~\bibnamefont {Martello}}, \bibinfo
  {editor} {\bibfnamefont {I.}~\bibnamefont {Osman}}, \bibinfo {editor}
  {\bibfnamefont {C.}~\bibnamefont {Roucairol}}, \ and\ \bibinfo {editor}
  {\bibfnamefont {S.}~\bibnamefont {Voss}},\ eds.,\ \href@noop {} {\emph
  {\bibinfo {title} {Meta-Heuristics: Advances and Trends in Local Search
  Paradigms for Optimization}}}\ (\bibinfo  {publisher} {Kluwer, Boston},\
  \bibinfo {year} {1999})\BibitemShut {NoStop}%
\bibitem [{\citenamefont {{Hartmann}}(2001)}]{hartmann:01b}%
  \BibitemOpen
  \bibfield  {author} {\bibinfo {author} {\bibfnamefont {A.~K.}\ \bibnamefont
  {{Hartmann}}},\ }\href@noop {} {\bibfield  {journal} {\bibinfo  {journal}
  {Phys. Rev. E}\ }\textbf {\bibinfo {volume} {63}} (\bibinfo {year}
  {2001})}\BibitemShut {NoStop}%
\bibitem [{\citenamefont {Palassini}\ and\ \citenamefont
  {Young}(1999)}]{Palassini99}%
  \BibitemOpen
  \bibfield  {author} {\bibinfo {author} {\bibfnamefont {M.}~\bibnamefont
  {Palassini}}\ and\ \bibinfo {author} {\bibfnamefont {A.~P.}\ \bibnamefont
  {Young}},\ }\href@noop {} {\bibfield  {journal} {\bibinfo  {journal} {Phys.
  Rev. Lett.}\ }\textbf {\bibinfo {volume} {83}},\ \bibinfo {pages} {5126}
  (\bibinfo {year} {1999})}\BibitemShut {NoStop}%
\bibitem [{\citenamefont {Pal}(2006)}]{Pal06b}%
  \BibitemOpen
  \bibfield  {author} {\bibinfo {author} {\bibfnamefont {K.~F.}\ \bibnamefont
  {Pal}},\ }\href@noop {} {\bibfield  {journal} {\bibinfo  {journal} {Physica
  A}\ }\textbf {\bibinfo {volume} {367}},\ \bibinfo {pages} {261} (\bibinfo
  {year} {2006})}\BibitemShut {NoStop}%
\bibitem [{\citenamefont {Boettcher}(2010{\natexlab{a}})}]{Boettcher10b}%
  \BibitemOpen
  \bibfield  {author} {\bibinfo {author} {\bibfnamefont {S.}~\bibnamefont
  {Boettcher}},\ }\href@noop {} {\bibfield  {journal} {\bibinfo  {journal} {J.
  Stat. Mech}\ ,\ \bibinfo {pages} {P07002}} (\bibinfo {year}
  {2010}{\natexlab{a}})}\BibitemShut {NoStop}%
\bibitem [{\citenamefont {Boettcher}(2005)}]{EOSK}%
  \BibitemOpen
  \bibfield  {author} {\bibinfo {author} {\bibfnamefont {S.}~\bibnamefont
  {Boettcher}},\ }\href@noop {} {\bibfield  {journal} {\bibinfo  {journal}
  {Eur.\ Phys.\ J.\ B}\ }\textbf {\bibinfo {volume} {46}},\ \bibinfo {pages}
  {501} (\bibinfo {year} {2005})}\BibitemShut {NoStop}%
\bibitem [{\citenamefont {Boettcher}\ and\ \citenamefont
  {Percus}(2000)}]{Boettcher00}%
  \BibitemOpen
  \bibfield  {author} {\bibinfo {author} {\bibfnamefont {S.}~\bibnamefont
  {Boettcher}}\ and\ \bibinfo {author} {\bibfnamefont {A.~G.}\ \bibnamefont
  {Percus}},\ }\href@noop {} {\bibfield  {journal} {\bibinfo  {journal}
  {Artificial Intelligence}\ }\textbf {\bibinfo {volume} {119}},\ \bibinfo
  {pages} {275} (\bibinfo {year} {2000})}\BibitemShut {NoStop}%
\bibitem [{\citenamefont {Boettcher}\ and\ \citenamefont
  {Percus}(2001)}]{Boettcher01a}%
  \BibitemOpen
  \bibfield  {author} {\bibinfo {author} {\bibfnamefont {S.}~\bibnamefont
  {Boettcher}}\ and\ \bibinfo {author} {\bibfnamefont {A.~G.}\ \bibnamefont
  {Percus}},\ }\href {\doibase 10.1103/PhysRevLett.86.5211} {\bibfield
  {journal} {\bibinfo  {journal} {Phys. Rev. Lett.}\ }\textbf {\bibinfo
  {volume} {86}},\ \bibinfo {pages} {5211} (\bibinfo {year}
  {2001})}\BibitemShut {NoStop}%
\bibitem [{\citenamefont {Middleton}(2004)}]{Middleton04}%
  \BibitemOpen
  \bibfield  {author} {\bibinfo {author} {\bibfnamefont {A.~A.}\ \bibnamefont
  {Middleton}},\ }\href@noop {} {\bibfield  {journal} {\bibinfo  {journal}
  {Phys. Rev. E}\ }\textbf {\bibinfo {volume} {69}},\ \bibinfo {pages}
  {055701(R)} (\bibinfo {year} {2004})}\BibitemShut {NoStop}%
\bibitem [{\citenamefont {{M\'ezard}}\ \emph {et~al.}(1987)\citenamefont
  {{M\'ezard}}, \citenamefont {Parisi},\ and\ \citenamefont {Virasoro}}]{MPV}%
  \BibitemOpen
  \bibfield  {author} {\bibinfo {author} {\bibfnamefont {M.}~\bibnamefont
  {{M\'ezard}}}, \bibinfo {author} {\bibfnamefont {G.}~\bibnamefont {Parisi}},
  \ and\ \bibinfo {author} {\bibfnamefont {M.~A.}\ \bibnamefont {Virasoro}},\
  }\href@noop {} {\emph {\bibinfo {title} {Spin glass theory and beyond}}}\
  (\bibinfo  {publisher} {World Scientific},\ \bibinfo {address} {Singapore},\
  \bibinfo {year} {1987})\BibitemShut {NoStop}%
\bibitem [{\citenamefont {Percus}\ \emph {et~al.}(2006)\citenamefont {Percus},
  \citenamefont {Istrate},\ and\ \citenamefont {Moore}}]{Percus06}%
  \BibitemOpen
  \bibfield  {author} {\bibinfo {author} {\bibfnamefont {A.}~\bibnamefont
  {Percus}}, \bibinfo {author} {\bibfnamefont {G.}~\bibnamefont {Istrate}}, \
  and\ \bibinfo {author} {\bibfnamefont {C.}~\bibnamefont {Moore}},\
  }\href@noop {} {\emph {\bibinfo {title} {Computational Complexity and
  Statistical Physics}}}\ (\bibinfo  {publisher} {Oxford University Press},\
  \bibinfo {address} {New York},\ \bibinfo {year} {2006})\BibitemShut {NoStop}%
\bibitem [{\citenamefont {{M{\'e}zard}}\ and\ \citenamefont
  {{Parisi}}(2001)}]{mezard:01}%
  \BibitemOpen
  \bibfield  {author} {\bibinfo {author} {\bibfnamefont {M.}~\bibnamefont
  {{M{\'e}zard}}}\ and\ \bibinfo {author} {\bibfnamefont {G.}~\bibnamefont
  {{Parisi}}},\ }\href@noop {} {\bibfield  {journal} {\bibinfo  {journal} {Eur.
  Phys. J. B}\ }\textbf {\bibinfo {volume} {20}},\ \bibinfo {pages} {217}
  (\bibinfo {year} {2001})}\BibitemShut {NoStop}%
\bibitem [{\citenamefont {Boettcher}(2003)}]{Boettcher03a}%
  \BibitemOpen
  \bibfield  {author} {\bibinfo {author} {\bibfnamefont {S.}~\bibnamefont
  {Boettcher}},\ }\href@noop {} {\bibfield  {journal} {\bibinfo  {journal}
  {Euro. Phys. J. B}\ }\textbf {\bibinfo {volume} {31}},\ \bibinfo {pages} {29}
  (\bibinfo {year} {2003})}\BibitemShut {NoStop}%
\bibitem [{\citenamefont {Sherrington}\ and\ \citenamefont
  {Kirkpatrick}(1975)}]{Sherrington75}%
  \BibitemOpen
  \bibfield  {author} {\bibinfo {author} {\bibfnamefont {D.}~\bibnamefont
  {Sherrington}}\ and\ \bibinfo {author} {\bibfnamefont {S.}~\bibnamefont
  {Kirkpatrick}},\ }\href@noop {} {\bibfield  {journal} {\bibinfo  {journal}
  {Phys. Rev. Lett.}\ }\textbf {\bibinfo {volume} {35}},\ \bibinfo {pages}
  {1792} (\bibinfo {year} {1975})}\BibitemShut {NoStop}%
\bibitem [{\citenamefont {Toulouse}(1977)}]{Toulouse77}%
  \BibitemOpen
  \bibfield  {author} {\bibinfo {author} {\bibfnamefont {G.}~\bibnamefont
  {Toulouse}},\ }\href@noop {} {\bibfield  {journal} {\bibinfo  {journal}
  {Communication on Physics}\ }\textbf {\bibinfo {volume} {2}},\ \bibinfo
  {pages} {115} (\bibinfo {year} {1977})}\BibitemShut {NoStop}%
\bibitem [{Note1()}]{Note1}%
  \BibitemOpen
  \bibinfo {note} {In the operations research literature, QUBO is usually
  defined as a maximization problem for $E$ without the sign; the conversion is
  trivial.}\BibitemShut {Stop}%
\bibitem [{\citenamefont {Beasley}()}]{Beasley98}%
  \BibitemOpen
  \bibfield  {author} {\bibinfo {author} {\bibfnamefont {J.~E.}\ \bibnamefont
  {Beasley}},\ }\href@noop {} {\enquote {\bibinfo {title} {Heuristic algorithms
  for the unconstrained binary quadratic programming problem},}\ }\bibinfo
  {note} {Tech. Rep., Management School, Imperial College (1998)}\BibitemShut
  {NoStop}%
\bibitem [{\citenamefont {Palassini}(2008)}]{Palassini08}%
  \BibitemOpen
  \bibfield  {author} {\bibinfo {author} {\bibfnamefont {M.}~\bibnamefont
  {Palassini}},\ }\href@noop {} {\bibfield  {journal} {\bibinfo  {journal} {J.
  Stat. Mech.}\ ,\ \bibinfo {pages} {P10005}} (\bibinfo {year}
  {2008})}\BibitemShut {NoStop}%
\bibitem [{\citenamefont {Gon{\c{c}}alves}\ and\ \citenamefont
  {Boettcher}(2008)}]{Goncalves08}%
  \BibitemOpen
  \bibfield  {author} {\bibinfo {author} {\bibfnamefont {B.}~\bibnamefont
  {Gon{\c{c}}alves}}\ and\ \bibinfo {author} {\bibfnamefont {S.}~\bibnamefont
  {Boettcher}},\ }\href@noop {} {\bibfield  {journal} {\bibinfo  {journal} {J.
  Stat. Mech.}\ ,\ \bibinfo {pages} {P01003}} (\bibinfo {year}
  {2008})}\BibitemShut {NoStop}%
\bibitem [{\citenamefont {Boettcher}(2010{\natexlab{b}})}]{Boettcher10a}%
  \BibitemOpen
  \bibfield  {author} {\bibinfo {author} {\bibfnamefont {S.}~\bibnamefont
  {Boettcher}},\ }\href@noop {} {\bibfield  {journal} {\bibinfo  {journal}
  {Euro. Phys. J. B}\ }\textbf {\bibinfo {volume} {74}},\ \bibinfo {pages} {363
  } (\bibinfo {year} {2010}{\natexlab{b}})}\BibitemShut {NoStop}%
\bibitem [{\citenamefont {Grest}\ \emph {et~al.}(1986)\citenamefont {Grest},
  \citenamefont {Soukoulis},\ and\ \citenamefont {Levin}}]{Grest86}%
  \BibitemOpen
  \bibfield  {author} {\bibinfo {author} {\bibfnamefont {G.~S.}\ \bibnamefont
  {Grest}}, \bibinfo {author} {\bibfnamefont {C.~M.}\ \bibnamefont
  {Soukoulis}}, \ and\ \bibinfo {author} {\bibfnamefont {K.}~\bibnamefont
  {Levin}},\ }\href@noop {} {\bibfield  {journal} {\bibinfo  {journal} {Phys.
  Rev. Lett.}\ }\textbf {\bibinfo {volume} {56}},\ \bibinfo {pages} {1148}
  (\bibinfo {year} {1986})}\BibitemShut {NoStop}%
\bibitem [{\citenamefont {Aspelmeier}\ \emph {et~al.}(2008)\citenamefont
  {Aspelmeier}, \citenamefont {Billoire}, \citenamefont {Marinari},\ and\
  \citenamefont {Moore}}]{Aspelmeier07}%
  \BibitemOpen
  \bibfield  {author} {\bibinfo {author} {\bibfnamefont {T.}~\bibnamefont
  {Aspelmeier}}, \bibinfo {author} {\bibfnamefont {A.}~\bibnamefont
  {Billoire}}, \bibinfo {author} {\bibfnamefont {E.}~\bibnamefont {Marinari}},
  \ and\ \bibinfo {author} {\bibfnamefont {M.~A.}\ \bibnamefont {Moore}},\
  }\href@noop {} {\bibfield  {journal} {\bibinfo  {journal} {Journal of Physics
  A: Mathematical and Theoretical}\ }\textbf {\bibinfo {volume} {41}},\
  \bibinfo {pages} {324008 (21pp)} (\bibinfo {year} {2008})}\BibitemShut
  {NoStop}%
\bibitem [{\citenamefont {Oppermann}\ \emph {et~al.}(2007)\citenamefont
  {Oppermann}, \citenamefont {Schmidt},\ and\ \citenamefont
  {Sherrington}}]{Oppermann07}%
  \BibitemOpen
  \bibfield  {author} {\bibinfo {author} {\bibfnamefont {R.}~\bibnamefont
  {Oppermann}}, \bibinfo {author} {\bibfnamefont {M.~J.}\ \bibnamefont
  {Schmidt}}, \ and\ \bibinfo {author} {\bibfnamefont {D.}~\bibnamefont
  {Sherrington}},\ }\href@noop {} {\bibfield  {journal} {\bibinfo  {journal}
  {Phys. Rev. Lett.}\ }\textbf {\bibinfo {volume} {98}},\ \bibinfo {pages}
  {127201} (\bibinfo {year} {2007})}\BibitemShut {NoStop}%
\bibitem [{\citenamefont {{Pankov}}(2006)}]{pankov:06}%
  \BibitemOpen
  \bibfield  {author} {\bibinfo {author} {\bibfnamefont {S.}~\bibnamefont
  {{Pankov}}},\ }\href@noop {} {\bibfield  {journal} {\bibinfo  {journal}
  {Phys. Rev. Lett.}\ }\textbf {\bibinfo {volume} {96}},\ \bibinfo {pages}
  {197204} (\bibinfo {year} {2006})}\BibitemShut {NoStop}%
\bibitem [{\citenamefont {Boettcher}\ and\ \citenamefont
  {Percus}(2004)}]{Boettcher04a}%
  \BibitemOpen
  \bibfield  {author} {\bibinfo {author} {\bibfnamefont {S.}~\bibnamefont
  {Boettcher}}\ and\ \bibinfo {author} {\bibfnamefont {A.~G.}\ \bibnamefont
  {Percus}},\ }\href@noop {} {\bibfield  {journal} {\bibinfo  {journal} {Phys.
  Rev. E}\ }\textbf {\bibinfo {volume} {69}},\ \bibinfo {pages} {066703}
  (\bibinfo {year} {2004})}\BibitemShut {NoStop}%
\bibitem [{\citenamefont {Boettcher}\ and\ \citenamefont
  {Grigni}(2002)}]{eo_jam}%
  \BibitemOpen
  \bibfield  {author} {\bibinfo {author} {\bibfnamefont {S.}~\bibnamefont
  {Boettcher}}\ and\ \bibinfo {author} {\bibfnamefont {M.}~\bibnamefont
  {Grigni}},\ }\href@noop {} {\bibfield  {journal} {\bibinfo  {journal} {J.
  Phys. A: Math. Gen.}\ }\textbf {\bibinfo {volume} {35}},\ \bibinfo {pages}
  {1109} (\bibinfo {year} {2002})}\BibitemShut {NoStop}%
\bibitem [{\citenamefont {Boettcher}(2015)}]{Boettcher2015}%
  \BibitemOpen
  \bibfield  {author} {\bibinfo {author} {\bibfnamefont {S.}~\bibnamefont
  {Boettcher}},\ }\href {\doibase 10.1016/j.phpro.2015.07.102} {\bibfield
  {journal} {\bibinfo  {journal} {Physics Procedia}\ }\textbf {\bibinfo
  {volume} {68}},\ \bibinfo {pages} {16} (\bibinfo {year} {2015})}\BibitemShut
  {NoStop}%
\bibitem [{Note2()}]{Note2}%
  \BibitemOpen
  \bibinfo {note} {We thank the referee for insisting on this
  consideration.}\BibitemShut {Stop}%
\bibitem [{\citenamefont {Beasley}(1996)}]{Beasley96}%
  \BibitemOpen
  \bibfield  {author} {\bibinfo {author} {\bibfnamefont {J.}~\bibnamefont
  {Beasley}},\ }\href@noop {} {\bibfield  {journal} {\bibinfo  {journal}
  {Journal of Global Optimization}\ }\textbf {\bibinfo {volume} {8}},\ \bibinfo
  {pages} {429} (\bibinfo {year} {1996})}\BibitemShut {NoStop}%
\bibitem [{\citenamefont {Billoire}(2006)}]{Billoire06}%
  \BibitemOpen
  \bibfield  {author} {\bibinfo {author} {\bibfnamefont {A.}~\bibnamefont
  {Billoire}},\ }\href {\doibase 10.1103/PhysRevB.73.132201} {\bibfield
  {journal} {\bibinfo  {journal} {Phys. Rev. B}\ }\textbf {\bibinfo {volume}
  {73}},\ \bibinfo {pages} {132201} (\bibinfo {year} {2006})}\BibitemShut
  {NoStop}%
\bibitem [{\citenamefont {de~Almeida}\ and\ \citenamefont
  {Thouless}(1978)}]{Almeida78}%
  \BibitemOpen
  \bibfield  {author} {\bibinfo {author} {\bibfnamefont {J.}~\bibnamefont
  {de~Almeida}}\ and\ \bibinfo {author} {\bibfnamefont {D.}~\bibnamefont
  {Thouless}},\ }\href@noop {} {\bibfield  {journal} {\bibinfo  {journal} {J.
  Phys. A}\ }\textbf {\bibinfo {volume} {11}},\ \bibinfo {pages} {983}
  (\bibinfo {year} {1978})}\BibitemShut {NoStop}%
\bibitem [{\citenamefont {J\"org}\ \emph {et~al.}(2008)\citenamefont {J\"org},
  \citenamefont {Katzgraber},\ and\ \citenamefont {Krzakala}}]{Joerg08}%
  \BibitemOpen
  \bibfield  {author} {\bibinfo {author} {\bibfnamefont {T.}~\bibnamefont
  {J\"org}}, \bibinfo {author} {\bibfnamefont {H.~G.}\ \bibnamefont
  {Katzgraber}}, \ and\ \bibinfo {author} {\bibfnamefont {F.}~\bibnamefont
  {Krzakala}},\ }\href {\doibase 10.1103/physrevlett.100.197202} {\bibfield
  {journal} {\bibinfo  {journal} {Physical Review Letters}\ }\textbf {\bibinfo
  {volume} {100}} (\bibinfo {year} {2008}),\
  10.1103/physrevlett.100.197202}\BibitemShut {NoStop}%
\bibitem [{\citenamefont {Larson}\ \emph {et~al.}(2013)\citenamefont {Larson},
  \citenamefont {Katzgraber}, \citenamefont {Moore},\ and\ \citenamefont
  {Young}}]{Larson13}%
  \BibitemOpen
  \bibfield  {author} {\bibinfo {author} {\bibfnamefont {D.}~\bibnamefont
  {Larson}}, \bibinfo {author} {\bibfnamefont {H.~G.}\ \bibnamefont
  {Katzgraber}}, \bibinfo {author} {\bibfnamefont {M.~A.}\ \bibnamefont
  {Moore}}, \ and\ \bibinfo {author} {\bibfnamefont {A.~P.}\ \bibnamefont
  {Young}},\ }\href {\doibase 10.1103/physrevb.87.024414} {\bibfield  {journal}
  {\bibinfo  {journal} {Physical Review B}\ }\textbf {\bibinfo {volume} {87}}
  (\bibinfo {year} {2013}),\ 10.1103/physrevb.87.024414}\BibitemShut {NoStop}%
\bibitem [{\citenamefont {Zintchenko}\ \emph {et~al.}(2015)\citenamefont
  {Zintchenko}, \citenamefont {Hastings},\ and\ \citenamefont
  {Troyer}}]{Zintchenko14}%
  \BibitemOpen
  \bibfield  {author} {\bibinfo {author} {\bibfnamefont {I.}~\bibnamefont
  {Zintchenko}}, \bibinfo {author} {\bibfnamefont {M.~B.}\ \bibnamefont
  {Hastings}}, \ and\ \bibinfo {author} {\bibfnamefont {M.}~\bibnamefont
  {Troyer}},\ }\href@noop {} {\bibfield  {journal} {\bibinfo  {journal} {Phys.
  Rev. B}\ }\textbf {\bibinfo {volume} {91}},\ \bibinfo {pages} {024201}
  (\bibinfo {year} {2015})}\BibitemShut {NoStop}%
\end{thebibliography}%

\end{document}